%% file: main.tex
\PassOptionsToPackage{twoside=false}{geometry}

\documentclass[acmsmall]{acmart}

\setcopyright{cc}
\setcctype{by}
\acmJournal{PACMHCI}
\acmYear{2026} \acmVolume{10} \acmNumber{5} \acmArticle{MHCI7668}
\acmMonth{8} \acmDOI{10.1145/3821671}

\usepackage{tikz}
\usepackage{amsmath}
\usepackage{multirow}
\usepackage{multicol}
\usepackage{enumitem}
\setitemize{topsep=2pt,parsep=0pt,partopsep=0pt,leftmargin=15pt}
\usepackage{graphicx}
\usepackage{filecontents}
\usepackage{xurl}
\usepackage{soul}
\usepackage{latexsym}
\usepackage{amsmath}
\usepackage{gensymb}
\usepackage{balance}
\usepackage{pifont}
\usepackage{anyfontsize}
\usepackage{xspace}
\usepackage{etoolbox}
\usepackage{subcaption}
\usepackage[inline]{./trackchanges}

\usepackage{atbegshi}
\AtBeginShipout{%
  \typeout{=== PAGE \thepage ===}%
  \typeout{textwidth=\the\textwidth}%
  \typeout{oddsidemargin=\the\oddsidemargin, evensidemargin=\the\evensidemargin}%
  \typeout{hoffset=\the\hoffset}%
  \typeout{marginparwidth=\the\marginparwidth, marginparsep=\the\marginparsep}%
}

\newcommand{\zerodisplayskips}{%
  \setlength{\abovedisplayskip}{3pt}%
  \setlength{\belowdisplayskip}{3pt}%
  \setlength{\abovedisplayshortskip}{3pt}%
  \setlength{\belowdisplayshortskip}{3pt}}
\appto{\normalsize}{\zerodisplayskips}
\appto{\small}{\zerodisplayskips}
\appto{\footnotesize}{\zerodisplayskips}

\AtBeginDocument{%
  \providecommand\BibTeX{{%
    \normalfont B\kern-0.5em{\scshape i\kern-0.25em b}\kern-0.8em\TeX}}}

\newcommand{\blue}[1]{{\color{black}#1}}
\newcommand{\newblue}[1]{{\color{black}#1}}

\usepackage{tabu}                      
\usepackage{booktabs}                  
\usepackage{lipsum}                    
\usepackage{mwe}                       
\usepackage [english]{babel}
\usepackage [autostyle, english = american]{csquotes}
\MakeOuterQuote{"}
\usepackage{comment}
\usepackage{multirow}

\newcommand{\SystemName}{HoloLens app\xspace}
\newcommand{\ProjectName}{ARTOO-DARTU\xspace}
\newcommand{\ObsPipeName}{ODM\xspace}

\begin{document}


\title[ARTOO-DARTU]{ARTOO-DARTU: Studying AR-HRC With AR Obstruction Mitigation During a Warehouse Task}

\author{Christian Fronk}
\orcid{0009-0000-3343-4882}
\affiliation{%
  \department{Electrical and Computer Engineering}
  \institution{Duke University}
  \city{Durham}
  \state{NC}
  \country{USA}
}
\email{christian.fronk@duke.edu}
\author{Hanting Ye}
\orcid{0000-0001-7306-5743}
\affiliation{%
  \department{Electrical and Computer Engineering}
  \institution{Duke University}
  \city{Durham}
  \state{NC}
  \country{USA}
}
\email{hanting.ye@duke.edu}
\author{Zhehan Qu}
\orcid{0009-0001-1997-3043}
\affiliation{%
  \department{Department of Computer Science}
  \institution{Duke University}
  \city{Durham}
  \state{NC}
  \country{USA}
}
\email{zhehan.qu@duke.edu}
\author{Maria Gorlatova}
\orcid{0000-0002-5477-7830}
\affiliation{%
  \department{Electrical and Computer Engineering}
  \institution{Duke University}
  \city{Durham}
  \state{NC}
  \country{USA}
}
\email{maria.gorlatova@duke.edu}

\renewcommand{\shortauthors}{Fronk et al.}


\begin{abstract}

Human-robot collaboration (HRC) often requires robot intentions and internal states to be conveyed to users for task efficiency and safety. Recently, augmented reality (AR) situated analytics provide such real-time robot feedback in HRC contexts. However, AR situated analytics can obstruct important real-world elements, posing safety and usability risks, especially when content is dynamically positioned relative to movements of mobile robots in a warehouse HRC scenario. In this paper, we introduce the \textbf{A}ugmented \textbf{R}eality \textbf{T}echnique \textbf{O}f \textbf{O}bstruction \textbf{D}eterrence while \textbf{A}iding \textbf{R}obotic \textbf{T}eaming for \textbf{U}sers (ARTOO-DARTU), an AR system tailored specifically for warehouse HRC that enables real-time robot situated analytics and control while preserving visibility of the real-world through an obstruction detection and mitigation pipeline (\ObsPipeName) that is uniquely suited for AR-HRC. \newblue{To evaluate ARTOO-DARTU, we developed Pocket MonstARs, a controlled gamified abstraction of HRC warehouse inventory picking in which virtual monsters serve as proxies for pick targets, while labeled and object-marked boxes preserve the real-world identification demands of the picking task.} In a 34-participant user study, we found that our designed AR situated analytics yielded a 46\% increase in efficiency on the overall HRC task, but only when the \ObsPipeName was active. Participants with the \ObsPipeName active were also 61\% faster on subtasks requiring visibility of the real world. Our findings demonstrate that, when paired with our developed \ObsPipeName to prevent real-world obstructions, the situated analytics in \ProjectName can significantly enhance efficiency and user experience in AR-HRC warehouse scenarios.
\end{abstract}






\begin{teaserfigure}
  \centering
    \includegraphics[width=\textwidth,height=0.35\textheight,keepaspectratio]{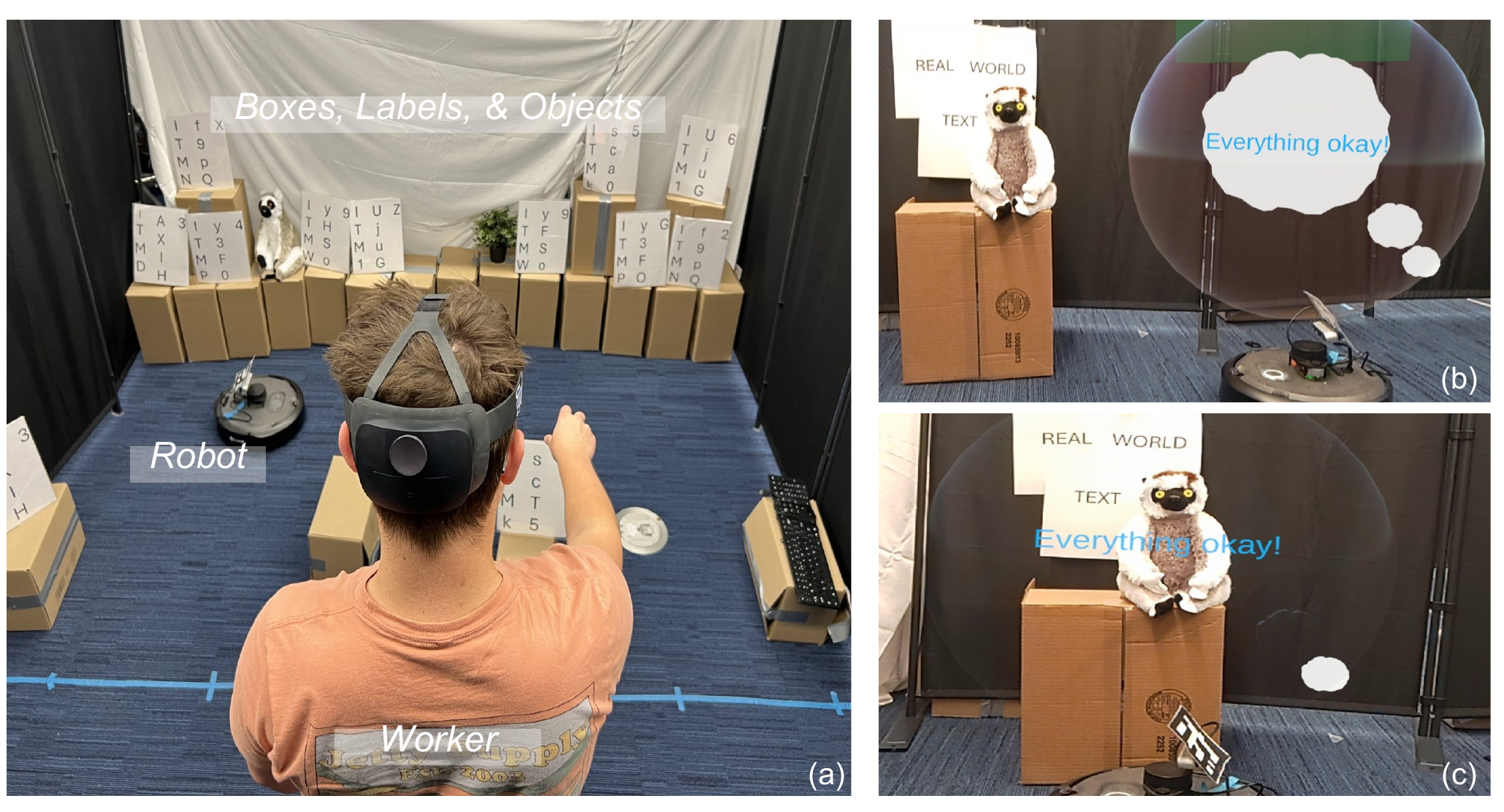}
  \caption{
    (a) The area allocated for our developed Pocket MonstARs task, where users collaborate with a robot to complete gamified picking targets represented as monsters. The task abstracts warehouse inventory picking by requiring users to identify target locations using physical labels, objects, and AR cues. Twelve labeled inventory boxes populate the 2.4$\times$2.4~m task area, along with four additional boxes containing YOLO-recognizable objects.
    (b) AR view showing robot ``thought bubble'' and location indicator situated analytics when real-world elements are unobstructed.
    (c) \blue{AR view after obstruction mitigation via the \ObsPipeName: thought bubble and location indicator become fully transparent when they overlap a real-world object and text.} 
  }
  \Description{
    Composite figure introducing the Pocket MonstARs study environment and ODM behavior. Panel (a) shows the physical Pocket MonstARs task area used in the user study. A participant wearing an AR headset points toward a small 2.4 by 2.4 meter area containing rows of cardboard boxes with printed labels and objects on top of some boxes. A robot is located in the task area. Panels (b) and (c) show first-person AR views of the robot near inventory boxes. In panel (b), the robot’s AR thought bubble is visible and readable because it does not block important real-world information. In panel (c), obstruction mitigation is active: the thought bubble and robot location indicator have become transparent where they would otherwise overlap real-world boxes and labels, allowing the physical objects and text to remain visible. The figure contrasts an unmitigated AR analytics display with an obstruction-mitigated display.
  }
  \label{fig:teaser}
\end{teaserfigure}

\maketitle

\input{./sec/introduction}
\input{./sec/relatedwork}

\input{./sec/case_study}
\input{./sec/system_design}

\input{./sec/game_design}

\input{./sec/study_design}
\input{./sec/study_results}
\input{./sec/discussion_future_work}

\input{./sec/conclusion}


\balance
\bibliographystyle{ACM-Reference-Format}
\bibliography{ref}

\end{document}

%% file: sec/introduction.tex
\section{Introduction} 
\label{sec:introduction}
Augmented reality (AR) can superimpose virtual content onto the real world to effectively guide a user in completing tasks, such as placing directional overlays for navigation~\cite{arbasicnavigation}. Beyond this, AR is also recognized for its potential in human-robot collaboration (HRC) scenarios~\cite{ismarhrc,staticvdynamicviz,robotinteractingvirtual,IkedaHRI2024ProgramAR,anticipatoryarm}, where it can enhance task efficiency by providing both human and robot with better awareness of each other's relative positions, objectives, and intentions. This mutual awareness is especially essential in some shared HRC environments like warehouses~\cite{warehouse}, where collaborative tasks often involve numerous mobile ground robots working alongside workers, such as in inventory picking (locating and retrieving items to fulfill orders)~\cite{Azadeh2017warehouserobotics,Guizzo2008warehouserobotskiva}.

\blue{Though valuable, improperly placed AR content can negatively impact worker efficiency because inventory picking depends on real-world visibility: workers must be able to see inventory labels and physical objects to complete their tasks.} These AR view management challenges are exacerbated in HRC scenarios, where virtual content is often displayed as if it is "attached" to a robotic collaborator, and so is subject to their movement and decisions. To assess how often AR-induced obstructions occur in an HRC warehouse environment, we developed and conducted a simulated case study set in an AR-HRC warehouse. The results indicate that obstructions caused by AR content are a significant concern in this context.

Motivated by this unique challenge of AR-HRC view management in dynamic and complex warehouse environments, this paper introduces the \textbf{A}ugmented \textbf{R}eality \textbf{T}echnique \textbf{O}f \textbf{O}bstruction \textbf{D}eterrence while \textbf{A}iding \textbf{R}obotic \textbf{T}eaming for \textbf{U}sers (ARTOO-DARTU), an AR system for the Microsoft HoloLens 2 that delivers detailed, real-time situated analytics about a mobile ground robot in a warehouse setting, and allows users to control this robot. The analytics of the system are designed to be informative and easily understood at a glance by a user involved in a warehouse task, with intuitive graphics and personable messages that portray a robot’s position, intent, and internal state. \newblue{To address the problem of obstruction caused by AR content, \ProjectName introduces \ObsPipeName, which, to our knowledge, is the first obstruction detection and mitigation pipeline designed specifically for robot-coupled situated analytics in AR-HRC.} \blue{Rather than assuming a known environment with predefined elements, the \ObsPipeName operates in environments with previously unseen real-world elements such as labels and objects. The \ObsPipeName detects obstructions and mitigates them while preserving the spatial coupling between analytics and the robot, ensuring real-world information remains visible without breaking the alignment of content with a robotic collaborator.} 



To investigate users' reactions to real-world obstructions caused by situated analytics in a warehouse-style task and evaluate \ProjectName, we develop Pocket MonstARs, \blue{a gamified task scenario that captures key aspects of warehouse inventory picking without directly replicating a full warehouse. \newblue{In the task, monsters serve as gamified representations of inventory picking targets, enabling a more engaging repeated-trial experience for participants while preserving the aspects of warehouse picking most relevant to AR obstruction: identifying targets using real-world cues, coordinating with a robotic collaborator, and maintaining awareness of both real and virtual elements within a structured evaluation setting.}} \newblue{Thus, Pocket MonstARs is intended as a controlled evaluation task for obstructions caused by AR content during warehouse AR-HRC, rather than a direct recreation of all warehouse operations.} We use Pocket MonstARs to evaluate \ProjectName and its \ObsPipeName through an IRB-approved $2\times3$ mixed factorial user study with 34 participants.

In summary, our main contributions are the following:

\begin{itemize}
  \item We created an AR system for HRC with mobile ground robots in warehouse settings, offering enhanced situated analytics and robot control. \textit{83\% of study participants preferred our solution over a baseline with no analytics.}
  \newblue{\item We introduce \textit{\ObsPipeName, an obstruction detection and mitigation pipeline for robot-coupled AR situated analytics in HRC}, and integrate it with our system providing AR-HRC analytics and control.}
  \item \blue{We created a \textit{controlled, gamified experimental task called Pocket MonstARs} \newblue{that uses monster targets as engaging proxies for inventory picking targets while preserving key warehouse-relevant demands, including both virtual and real-world visibility.} This enables systematic evaluation of AR analytics and obstruction.} \textit{We found that real-world subtasks were completed up to 46\% faster with analytics, but only when the \ObsPipeName was active. Participants with the \ObsPipeName were 39\% faster overall and 61\% faster on real-world subtasks.}
\end{itemize}

The remainder of this paper is organized as follows: Sec.~\ref{sec:relatedwork} reviews related work, followed by the motivating case study in Sec.~\ref{sec:motivation}. We present \ProjectName's system design in Sec.~\ref{sec:systemdesign} and introduce Pocket MonstARs in Sec.~\ref{sec:gamedesignsec}. Sec.~\ref{sec:userstudydesign} outlines our user study, and the study results are presented in Sec.~\ref{sec:studyeval}. We discuss our findings and future work in Sec.~\ref{sec:limitationsfw}. We then conclude the paper in Sec.~\ref{sec:conclusion}.


%% file: sec/relatedwork.tex
\section{Related Work}
\label{sec:relatedwork}

\noindent\textbf{AR Situated Analytics for HRC.}
AR situated analytics have been widely used in diverse scenarios~\cite{arnavigation,arnavuncertainty,sitelens,pearl,relive,ragrug,sitar}, including HRC, and are shown to increase task efficiency~\cite{ismarhrc,staticvdynamicviz,robotinteractingvirtual,IkedaHRI2024ProgramAR}, enhance robot control~\cite{ChanHRI2022HumanRobotArmManufacturingCollab,HRI_2019_Drone_Teleop_Indoors,HRI_2018_Teleop_Indoors_Analytics_Comparison}, and improve user understanding of robot state especially during delays~\cite{cozmo} or unexpected behavior~\cite{anticipatoryarm}, e.g., to determine whether the robot stopped due to a sensor failure or a depleted battery. However, due to the high mobility of robots in warehouse environments, situated analytics provided during HRC can detrimentally affect workers’ perception of critical elements, such as object labels and safety signs. To investigate this important AR safety concern, we built the first simulated warehouse environment designed to explore the degree at which HRC situated analytics can impact a worker's view in a warehouse scenario, as presented in Sec.~\ref{sec:motivation}. We developed ARTOO-DARTU, an AR-HRC system for mobile warehouse robots that delivers situated analytics, enables robot control, and mitigates real-world occlusions through its \ObsPipeName. \blue{We evaluate \ProjectName in a gamified experimental task in an emulated warehouse setting that captures key perceptual and coordination demands of warehouse HRC, assessing its effectiveness in supporting task performance and its ability to reduce visual obstructions.}

\noindent\textbf{AR View Management.}
Improper AR view management can introduce detrimental effects, as AR content may render real-world objects unrecognizable. Cheng et al.~\cite{perceptionmanip} show that intentionally overlaying virtual content of a contrasting color on real-world elements can limit a user's real-world awareness. Furthermore, Sajid et al.~\cite{sajid2025juststopdoingnow} show that users struggle to recognize when AR content is detrimental and are often unsure how to respond during remote AR collaboration tasks. Existing AR view management techniques attempt to mitigate such issues by moving content to the ceiling, as in Satkowski et al.~\cite{Satkowski2022CeilingorFloorPlacement}, or optimizing the AR content placement for stationary environmental elements, such as when labeling buildings~\cite{bellviewmanagementuist2001,maassviewmanagementsmartgraphics2006,grassetviewmanagementismar2012}.
However, in the AR-HRC warehouse task we address, AR content used for situated analytics is often tied to the robot's real-world position, and therefore cannot be moved arbitrarily without breaking spatial alignment, nor should the robot be forced to move. Moreover, the unpredictable movement of both users and robots, combined with the dynamic nature of warehouses where elements like boxes are frequently relocated, renders methods that depend on fixed environment elements unsuitable. These challenges necessitate an AR view management approach for AR-HRC, which we are the first to develop, that automatically controls virtual content to maintain real-world clarity in dynamic warehouse environments with mobile elements.

\noindent\textbf{AR Obstruction Detection.}
A key challenge in AR view management is handling obstructions of the user's view of the real-world, for which many recent works have proposed remedies for. Many prior methods~\cite{Davari2020OcclusionManGlanceable,Lebeck2017SecureOutputinSim,ShahICIDT2012OcclusioninAR} rely on predefined models of the world and objects, and therefore do not generalize to environments with mobile elements as is required for an HRC scenario. Recently, Xiu et al.~\cite{xiu2025viddar} proposed detecting real-world AR obstructions using a vision-language model. However, their system requires calling an API to access closed-source vision-language models hosted on cloud servers, which introduces the risk of sensitive information leakage and leads to high latency; these factors make approaches like this unsuitable for deployment in real-world AR-HRC warehouse environments, as these environments may contain sensitive industry information, and obstructions must be dealt with quickly to ensure safety and efficiency. ARTOO-DARTU's \ObsPipeName is designed for dynamic, information-rich warehouse environments and operates in near real-time without requiring cloud infrastructure. \blue{To evaluate its effectiveness, we conduct a user study centered on a gamified task that captures key characteristics of warehouse HRC. Unlike prior work~\cite{Davari2020OcclusionManGlanceable,Lebeck2017SecureOutputinSim,ShahICIDT2012OcclusioninAR,xiu2025viddar}, our approach enables systematic measurement of both task performance and user experience.}


%% file: sec/case_study.tex
\section{AR-HRC Warehouse Simulated Case Study}
\label{sec:motivation}

\begin{figure}[t]
\centering
\includegraphics[width=0.7\columnwidth]{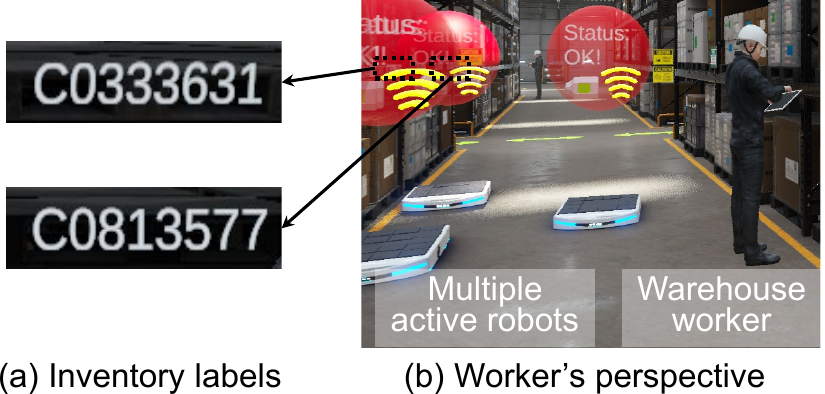}
\caption{The simulated warehouse environment. (a) Zoomed-in views of inventory labels. (b) Worker’s perspective with multiple active ground robots, and a coworker. 
}
\Description{
Composite figure showing the simulated warehouse environment used to study AR-induced obstructions. Panel (b) shows zoomed-in examples of inventory labels, consisting of white text on black label backgrounds with alphanumeric codes. Panel (b) presents the worker’s perspective inside a warehouse aisle. Multiple mobile ground robots are visible on the floor, each with large red AR status bubbles above them. Shelving and inventory areas line the aisle, and a warehouse worker is visible farther down the aisle. The view illustrates a visually dense environment where AR robot analytics could block safety-relevant or task-relevant real-world information. Together, the panels emphasize that inventory labels are important real-world warehouse elements that AR content should avoid obstructing.
}
\label{motivationalfigure}
\end{figure}

As a motivating case study on the frequency of AR-induced obstructions in a warehouse, we developed a virtual simulation of a near-future HRC warehouse scenario within the Unity engine, as seen in Fig.~\ref{motivationalfigure}(b). We design the layout of this environment and the arrangement of elements within it based on a typical warehouse design as in~\cite{Azadeh2017warehouserobotics}, which includes two types of warehouse systems: a compact storage system composed of stored boxes with labels, and a retrieval and fulfillment system involving workers and robots. The key elements of the environment are as follows: multiple shelves forming lanes for storing and retrieving boxes; boxes placed on the shelves containing goods for distribution; labeled shelf positions which must remain visible to ensure efficiency; and a mobile worker wearing an AR headset who performs an inventory picking task alongside collaborative mobile robots, guided by situated analytics displayed above the robots.

\noindent\textbf{Environment and Object Initialization.}
The warehouse environment, sourced from the Unity Asset Store~\cite{UnityWarehouse_2024}, measures 37.5×8.5×64.7~m and is rendered using Unity’s high-definition render pipeline. During the simulation, a simulated worker traverses 2 central lanes formed by 4 shelves, stocking labeled locations on the 2 innermost shelves with their corresponding incoming packages. 

\noindent\textbf{Mobile Worker and Robots.}
The worker navigates at 2~m/s using Unity’s NavMesh~\cite{UnityNavMesh}, placing boxes at randomized labeled target shelf locations (24 total) before returning to a pickup point for the next box. Ground robots move at 1~m/s between six randomly selected destinations along the lanes, occasionally pausing (0.1\% chance per frame) for 1~s to simulate box inspection. Each robot carries AR analytics above it, including a red 1~m radius status sphere with text (0.63×0.5~m) and a Wi-Fi symbol (0.04×0.3×0.38~m).

\noindent\textbf{Experiments.}
To quantify the extent to which AR-HRC analytics might obstruct visibility in a warehouse environment  and provide guidance for the design of the \ObsPipeName, we conducted 5 experiments varying the number of active robots from 2 to 10 in increments of 2, with 10 trials per setting. This progressive increase in the number of active robots allowed us to examine how obstruction from AR content scales with an increasing number of robots. To expedite the experiments, we set Unity’s timescale to 5×, reducing the average trial duration to 91.2~s. During each trial, we captured 2 types of obstruction data:  \textit{“Target Anywhere” (TA)} and \textit{"Target in FOV" (TFOV)}. TA measured whether each current label target was obstructed in any frame, regardless of whether it was currently in the worker’s field of view. TFOV counted only obstructions of elements within the camera frustum. For both measures, obstruction was determined using a raycast from the worker’s head, and we averaged the percentage of obstructed frames across trials for each robot-count condition.


\noindent\textbf{Results.}
Our results show with only 2 robots, label obstructions are relatively infrequent, averaging around 0.1\% in the TA and TFOV scenarios. However, as the number of robots increases, the frequency of obstructions also increases dramatically. With 10 robots, target labels are obstructed for 0.8\% of the frames on average in the TA scenario and for 0.4\% of the frames in the TFOV scenario. This trend shows that AR content can frequently cause detrimental obstructions of critical warehouse elements, especially as more robots become active, and illustrates the need for AR obstruction management methods in AR-assisted HRC tasks. We draw from these results and the parameters of the warehouse environment to create our study and the \ObsPipeName.


%% file: sec/system_design.tex
\section{System Design}
\label{sec:systemdesign}
In this section, we provide an overview of \ProjectName, a system that provides robot situated analytics and robot control in an AR-assisted HRC warehouse scenario while counteracting real-world AR obstructions. Given the dynamic nature of warehouse tasks, particularly when combined with mobile robotic collaborators, and the criticality of real-world visibility, \ProjectName is designed with several priorities in mind:
\begin{itemize}
\item[1.] Situated analytics that convey key aspects of a mobile robot that are easily interpretable when users are engaged in a warehouse task.
\item[2.] Convenient AR control of a robotic collaborator requiring minimal interface additions. 
\item[3.] Mitigation of AR obstructions of the real-world in a dynamic environment with mobile elements, without mitigation breaking spatial alignment of situated analytics. 
\end{itemize}
\blue{We do not position ARTOO-DARTU as a system for responding to warehouse accidents after they occur. Rather, our intended application is routine AR-assisted warehouse HRC, such as inventory picking, where workers must interpret robot state while maintaining visibility of task-relevant physical information. Thus, our focus is preventative: preserving real-world visibility while still providing robot state and intent information through AR situated analytics.}

\subsection{System Implementation}
\begin{figure}[t]
  \centering
  \includegraphics[width=.9\columnwidth]{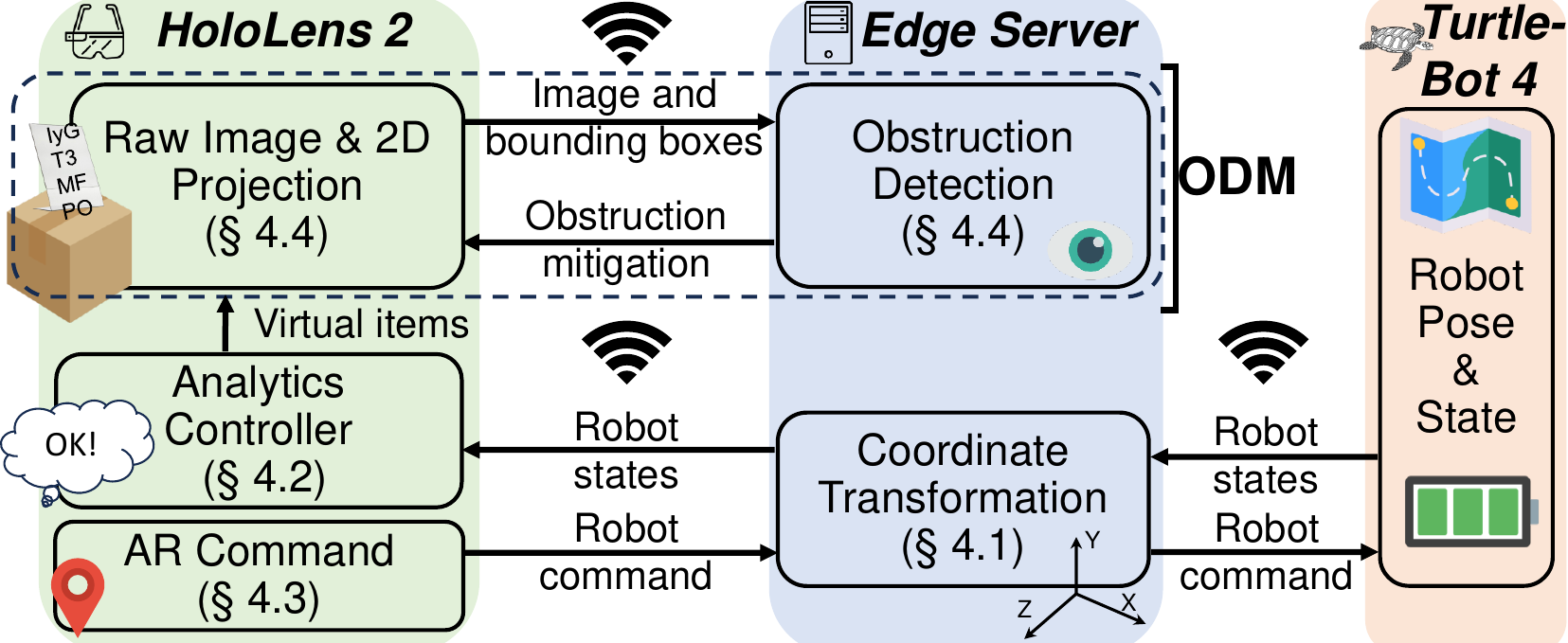}
  \caption{\ProjectName system diagram.}
  \Description{System diagram of ARTOO-DARTU showing communication among three main components: the HoloLens 2 headset, an edge server, and the TurtleBot 4 robot. On the left, the HoloLens 2 contains the AR command interface, coordinate transformation module, analytics controller, and virtual AR items. The analytics controller renders robot-related AR content such as a thought bubble with “OK!” and coordinate axes labeled X, Y, and Z. The HoloLens sends robot commands to the edge server, receives robot states, and sends raw images plus 2D projections for obstruction detection. In the center, the edge server receives robot commands, robot pose and state information, images, and bounding boxes. It performs coordinate transformation and obstruction detection as part of the ODM pipeline. On the right, the TurtleBot 4 exchanges robot commands and robot state information with the edge server. A feedback arrow from the server to the headset indicates obstruction mitigation information being returned so the headset can adjust virtual items. The diagram communicates that the headset handles AR display and interaction, the robot handles navigation and state reporting, and the edge server mediates communication while running computationally demanding obstruction detection.}
  \label{archdiagram}
\end{figure}

\ProjectName consists of 3 components: a HoloLens 2 headset, a TurtleBot 4, and an edge server. The headset runs a Unity 2022.3.10f1 application developed using Microsoft's Mixed Reality Toolkit~\cite{MRTK} and displays situated analytics and enables user control of the robot. The TurtleBot 4 runs ROS 2 Humble~\cite{roshumble} and executes both preinstalled ROS nodes and custom nodes that we created. The edge server, equipped with an NVIDIA RTX 3060 GPU, hosts a FastAPI-based~\cite{fastapi} Python API we developed to facilitate communication between headset and robot. The server also serves as a local compute resource for offloading more demanding tasks, such as running the character recognition and object detection models used in the \ObsPipeName. To align the situated analytics with the TurtleBot and enable robot control, \ProjectName computes a one-time transformation between the TurtleBot map frame and the \SystemName world frame using ArUco marker observations from both systems, as is common in prior work~\cite{Chen2021PinpointFlyCHI}, and the Kabsch–Umeyama algorithm~\cite{kabschumeyama}. The overall architecture of \ProjectName is shown in Fig.~\ref{archdiagram}. Sec.~\ref{sec:visualizationapplication} explains the situated analytics we developed, and how the \SystemName analytics controller works with the server and robot to enable these features (\textit{Priority 1}). Sec.~\ref{sec:robotcontrol} outlines the robot control method we developed (\textit{Priority 2}). Finally, Sec.~\ref{sec:obspipedesign} outlines how the \ObsPipeName prevents obstructions of real-world elements by virtual content (\textit{Priority 3}). 

\subsection{Robot Situated Analytics}
\label{sec:visualizationapplication}
\ProjectName provides analytics indicators in the \SystemName designed to display TurtleBot information most relevant to a human warehouse collaborator: position, internal state, command progress, and intended motion. \blue{We selected these categories because prior AR-HRC work has shown that analytics displaying these characteristics in particular can improve task efficiency, robot control, and user understanding of robot state~\cite{ismarhrc,staticvdynamicviz,robotinteractingvirtual,IkedaHRI2024ProgramAR,anticipatoryarm}. Our visualizations focus on information that helps users understand where the robot is, what it is doing, whether it requires attention, and how it will move next.} Thus we  design 3 main status indicators: a \textit{robot location indicator}, a \textit{thought bubble}, and a \textit{command status screen}. We also design 4 \textit{mini indicators} that give visual alerts regarding the TurtleBot's internal state, as well as \textit{trajectory indicators} that show the robot's intended movement path.

\begin{figure*}[t]
  \centering
  \includegraphics[width=\columnwidth]{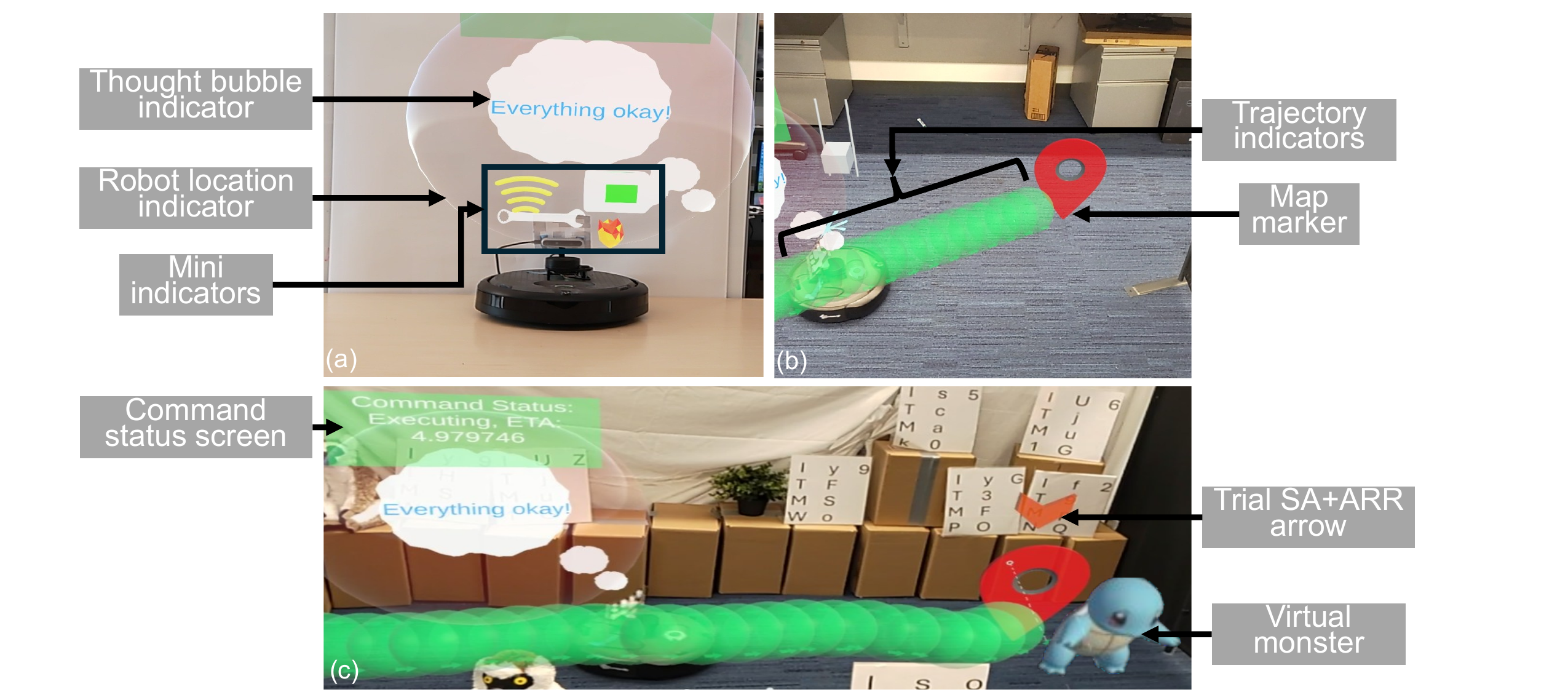}
  \caption{The setup and AR view of \ProjectName. (a) Situated analytics indicators for the robot. (b) Use of the map marker to command the robot and the trajectory indicators. (c) Commanding the robot to the location of a virtual monster during Pocket MonstARs Situated Analytics Trial with Arrows (SA+ARR). An arrow highlights the current box target location. \blue{Note \textit{the \ObsPipeName is inactive here for analytics clarity.}}}
  \Description{Composite figure showing ARTOO-DARTU’s situated analytics and robot-control interface in the physical Pocket MonstARs setup. Panel (a) shows a TurtleBot with several AR indicators placed around it. Labels point to a thought bubble indicator above the robot, a spherical robot location indicator surrounding the robot, mini indicators showing robot status information, and a command status screen. The thought bubble contains a short robot message, while the mini indicators include status icons such as Wi-Fi and battery symbols. Panel (b) shows the AR map marker used to command the robot and a series of green trajectory indicators on the floor, forming a path from the robot toward the marker. Panel (c) shows a wider AR task view during a Situated Analytics with Arrows trial. A green trajectory path extends across the floor toward a virtual monster, while a red map marker is placed near the monster’s location. An orange arrow highlights a box target location, and the robot analytics remain visible near the robot. The figure illustrates how ARTOO-DARTU combines robot state information, planned motion visualization, direct robot control, and optional target arrows in the same AR workspace}
  \label{fig:indicators}
\end{figure*}

\noindent\textbf{Main Status Indicators.}
The \SystemName features 3 primary situated analytics indicators:
\begin{itemize} 
\item \emph{Robot location indicator:} A spherical robot location indicator with a 65~cm diameter that indicates the TurtleBot's position, and encompasses all the following indicators.
\item \emph{Thought bubble indicator:} A 35x4x32~cm thought bubble that displays personified messages about the robot's entire internal state "from" the TurtleBot.
\item \emph{Command status screen:} A 46x1x20~cm command status screen that notifies the user when a command is received and how long it will take to complete the command.
\end{itemize}
\blue{Each indicator presents a single type of information to support clarity and glanceability, reflecting prior work showing the benefits of quickly accessible AR information~\cite{Davari_2022_VR_JustifyGlanceableAR}. We use a spherical indicator because it remains legible from different viewpoints and serves as a persistent anchor for robot information. Placing the other indicators within it keeps information visually grouped with the  robot rather than distributed across unrelated parts of the user view.} Previous work finds adding personification to robot communication is effective in enhancing trust and is preferred by users~\cite{SongVR2025VirtualUltrasound} and \blue{that communication is valuable during unexpected behavior~\cite{cozmo}}; with this in mind, we design our system such that personified messages "from" the TurtleBot appear within the thought bubble when a change in the internal state is detected by the TurtleBot and then received by the \SystemName. For example, when the TurtleBot's battery drops from 100\% to 75\% a message reading "I'm a little tired!" displays in the thought bubble. We choose a thought bubble to present this information to convey the impression that they represent the robot’s internal thoughts, leveraging a visual metaphor commonly found in cartoons, thus supporting intuitive user interpretation. \blue{The command status screen provides explicit feedback that a user command has been received and is being executed, closing the interaction loop after AR-based robot control. We include estimated completion time because this feedback helps users distinguish between a robot that is still executing a command and one that requires renewed attention.}

\noindent\textbf{Mini Indicators.}
To further support understanding of the TurtleBot’s internal state, 4 additional mini indicators appear within the location indicator when relevant. Each visually represents a specific aspect of robot internal state, with designs chosen to clearly \blue{reflect their semantic meanings as suggested in prior work~\cite{isherwood2009graphics_JustifySemanticIconDesign}}. These include: a 5×9×7~cm fire icon for heat level, a 16×10×10~cm battery for charge level, a 12×10×1~cm Wi-Fi symbol for network conditions, and an 18×5×1~cm wrench indicating maintenance needs. These indicators, along with the main status indicators, are shown in Fig.~\ref{fig:indicators}(a).

\noindent\textbf{Trajectory Indicators.}
To help users anticipate the TurtleBot's intended movement in a potentially crowded warehouse environment, the \SystemName visualizes the robot's planned trajectory as a sequence of green waypoint spheres with directional arrows, as shown in Fig.~\ref{fig:indicators}(b) and (c). This design supports robot intent awareness by communicating the spatial route the robot will follow and its intended heading at each point, as prior work has shown that similar visualizations can improve users' understanding of robot motion intent~\cite{modifyintent}. \blue{We use discrete,  sampled indicators rather than a continuous opaque path to limit visual clutter and display the indicators at floor-level to preserve visibility of the physical environment. The \SystemName also displays only the most recent planned trajectory, overwriting previous paths whenever a new destination is assigned, which prevents outdated indicators from accumulating.} When the TurtleBot is assigned a destination, it sends the server API its path as an ordered array of TurtleBot map-frame coordinates, along with the robot's ETA and execution status, as shown in Fig.~\ref{archdiagram}. The server API transforms each path point into \SystemName's coordinate frame, and the \SystemName renders the updated waypoints and arrows as they are received.

\subsection{AR Robot Control}
\label{sec:robotcontrol}

The \SystemName enables TurtleBot control through a virtual map marker, as shown in Fig.~\ref{fig:indicators}(b) and (c). Using MRTK’s far interaction gesture, the user can drag the marker to a location and release it, sending a destination to the server API as seen in Fig.~\ref{archdiagram}. This drag-and-drop control style is intended to let users direct the TurtleBot without diverting attention from their physical workspace. The approach is informed by prior work that demonstrated the effectiveness of similar methods for robot control~\cite{Chen2021PinpointFlyCHI}. We use a map marker as the visual metaphor for the control interface, drawing on its familiarity from common navigation and mapping applications.



\subsection{Obstruction Detection and Mitigation (\ObsPipeName)}
\label{sec:obspipedesign}
\begin{figure}[t]
  \centering
  \includegraphics[width=.9\columnwidth]{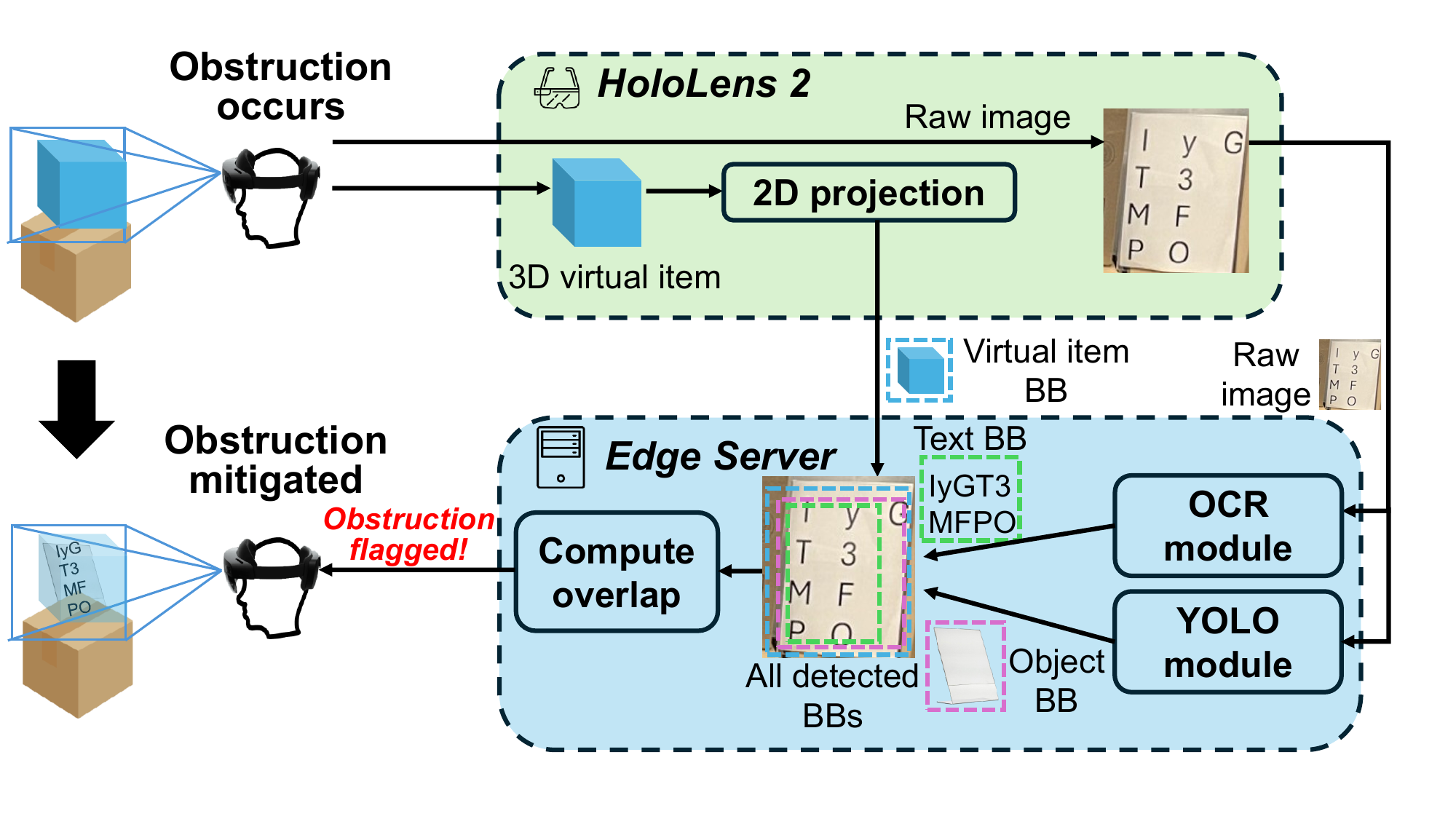}
  \caption{The flow of the \ObsPipeName for a case in which an AR item obstructs a real-world labeled inventory box.}
  \Description{Workflow diagram of the obstruction detection and mitigation pipeline, or ODM, for a case where an AR item blocks a real-world labeled inventory box. On the left, a real-world labeled box is shown with a blue virtual item in front of it; a HoloLens user views the scene, and the diagram labels this as an obstruction occurring. The HoloLens 2 section shows two parallel pieces of information: a 3D virtual item is projected into a 2D bounding box, and a raw camera image of the real world is captured. The raw image contains the box and label with text, while the virtual item bounding box marks where AR content would appear in the user’s view. These data are sent to the edge server. Inside the edge server, OCR detects text bounding boxes, YOLO detects object bounding boxes, and all detected bounding boxes are combined with the virtual item bounding box. A “compute overlap” step compares the projected AR item with the detected real-world objects and text. If the overlap is large enough, the system flags an obstruction. The diagram then shows the obstruction mitigated: the AR item becomes transparent or visually removed from the obstructing region, allowing the labeled box to be visible again. The figure communicates that ODM detects obstruction by comparing projected AR content against detected real-world text and object regions in the camera image.}
  \label{obspipediagram}
\end{figure}

\noindent
The goal of the \ObsPipeName is to preserve visibility of task-relevant real-world elements while maintaining the spatial coupling between AR analytics and a mobile robotic collaborator. \newblue{We frame obstruction mitigation in AR-HRC as a system-level view-management problem with three requirements that are not jointly addressed by prior AR obstruction approaches:} (1) the obstructing virtual content is dynamically coupled to a moving robot \newblue{and therefore cannot be freely relocated without weakening its communicative meaning}; (2) the real-world elements that must remain visible, such as labels and objects, may be previously unseen and may change over time; and \newblue{(3) mitigation must occur quickly enough to support ongoing collaboration without interrupting the robot's task execution.} \newblue{The \ObsPipeName addresses this problem through a projection-based pipeline that detects overlaps between robot-coupled AR analytics and task-relevant real-world text or objects, then mitigates the obstructing virtual elements while preserving their spatial association with the robot.} Thus, the contribution of the \ObsPipeName is not a new object detector, OCR model, or projection algorithm, but an AR-HRC-specific obstruction mitigation pipeline that \newblue{operationalizes these constraints in an implemented system and is evaluated in a controlled warehouse-style HRC task.} The overall \ObsPipeName pipeline is shown in Fig.~\ref{obspipediagram}.


\noindent\textbf{Raw Image and 2D Projections.}
To detect obstructions from AR content without disrupting user experience or incurring large data transmission, the \SystemName employs a projection-based method that avoids direct comparison between augmented and non-augmented views. While directly comparing these views would be a straightforward approach, the HoloLens 2 cannot capture both simultaneously, and doing so would introduce additional latency and bandwidth overhead. \blue{More importantly, in our setting the real-world elements that must remain visible are \textit{not known a priori} and are not available as 3D models, as would be the case in real dynamic warehouses. As such, simple geometric approaches (e.g., raycasting between known objects) are insufficient, as the system must reason about potential occlusions with \textit{previously unseen} real-world content. To address this, we adopt a projection-based formulation that enables comparison between virtual content and detected real-world elements.}

Specifically, the \SystemName captures images from the HoloLens’s front camera using the Vuforia Unity package; these images reflect the user’s view of the real world without any rendered virtual content. For each visible virtual item, a reference to the item is stored within the image and projections module, and the system computes a 2D projection bounding box onto the captured image using the intrinsic matrix of the main Unity camera and the item’s 3D rendering bounds. This results in an array of axis-aligned bounding boxes, each paired with the corresponding virtual item. These boxes indicate where virtual content would appear within the user’s real-world view, enabling subsequent overlap reasoning with detected real-world elements. The image is then downsampled to 25\% resolution, and both the image and projected bounding boxes are sent to the server API via a POST request.

\noindent\textbf{Obstruction Detection.}
The server API processes the received image using the well-tested YOLOv5~\cite{yolov5} model for object detection and the EasyOCR~\cite{EasyOCR} model for text detection, both returning bounding boxes for all detected objects and text. \blue{We deliberately build the ODM from well-established perception components rather than proposing a new detector or OCR model. This design choice reflects the focus of our contribution: identifying the AR-HRC obstruction problem and developing a system-level method that can mitigate when robot-coupled AR analytics obstruct task-relevant real-world content. Detections are performed without prior knowledge of the identity, location, or appearance of real-world elements so the system can reason about obstructions involving previously unseen objects and text, which is critical in warehouse environments where layouts may change over time.} For text detection, the server does not consider what detected text actually says, as we consider all real-world text to be of high importance. \blue{Text detection here focuses on inventory box labels, which are central to warehouse picking tasks. These labels must remain readable for users to correctly identify target items, making them a critical class of real-world elements to preserve. While safety signage is also important, inventory labels represent the primary task-relevant text in picking workflows; we therefore use OCR-based detection to localize such labels, enabling the approach to generalize to different environments.}

All object and text bounding boxes are placed as axis-aligned rectangles onto the image, sharing the same coordinate system as the projection boxes. \blue{For each virtual item, the system computes its overlap with all detected object and text bounding boxes. This enables reasoning over multiple virtual elements and multiple detected real-world elements simultaneously, rather than relying on geometric tests between known objects.} The overlap between the bounding box of each virtual item with each object and text bounding box is then computed. If any virtual item's bounding box overlaps more than 30\% of one of the object or text boxes, the name of that item is flagged and stored. We choose a 30\% threshold empirically, as we observe that label readability of the nature found in our task is often compromised beyond this level of occlusion. After all comparisons, the flagged virtual item names are returned to the \SystemName, and the app is notified to begin mitigation.

\noindent\textbf{Obstruction Mitigation.}
To mitigate obstructions, \blue{the \SystemName forces all flagged items to transition to being fully transparent when an obstruction is detected, regardless of whether they are initially rendered as opaque or semi-transparent, as in Fig.~\ref{fig:teaser}(c)}. This ensures that underlying real-world elements remain clearly visible, as even semi-transparent content can obscure fine details. \blue{In AR-HRC scenarios the virtual content is spatially coupled to a moving robot and conveys task-relevant information about its state and intent; as such, our mitigation operates purely through visual modulation, preserving spatial alignment while restoring visibility of real-world elements.} This approach balances obstruction management with spatial alignment and usability of the analytics in a mobile HRC scenario, as the \ObsPipeName does not require the robot or virtual items to move. This minimizes task disruption, maintaining the intended positioning of the analytics with the robot and preventing interruption of the robot’s goals. \blue{The system can also handle complex scenes in which several virtual indicators may simultaneously obstruct different task-relevant objects or text.}

Each item remains fully transparent until it has moved 20~cm or more away from the location at which the obstruction was detected in 3D space. We measure this Euclidean distance in the \SystemName subject to Unity's coordinate space. We choose 20~cm as we find experimentally that visibility of objects and labels typically returns once the content has moved this distance. \blue{This distance-based hysteresis prevents rapid toggling of visibility during robot motion.} After mitigation, the process repeats. Our experiments show that \ObsPipeName completes end-to-end processing in 346~ms on average, supporting near real-time obstruction detection and mitigation.

%% file: sec/game_design.tex
\vspace{-0.35cm}
\section{Pocket MonstARs Game Design}
\label{sec:gamedesignsec}
\blue{To evaluate \ProjectName in a controlled yet representative setting, we developed Pocket MonstARs (PM), a gamified task scenario that abstracts key components of a HRC warehouse inventory picking task. Pocket MonstARs is designed as a smaller, controlled, gamified abstraction of a warehouse picking scenario rather than a literal reproduction of a full warehouse. It preserves the key demands needed to study warehouse HRC in a repeatable manner: robot coordination, AR analytics interpretation, real-world target identification, and maintaining awareness of both virtual and physical information. To clarify how the game represents a warehouse HRC scenario, we explicitly map game elements to their real-world counterparts:} 
\blue{
\begin{itemize}
    \item Inventory boxes $\rightarrow$ warehouse storage locations
    \item Box labels and objects $\rightarrow$ real-world identification cues used in picking tasks
    \newblue{\item Monsters $\rightarrow$ gamified pick targets requiring robot interaction}
    \item Instruction prompts $\rightarrow$ task prompts indicating where workers should send the robot to "pick" target item 
\end{itemize}
This mapping allows the game to preserve the perceptual and coordination challenges of warehouse picking while enabling repeatable evaluation. Importantly, the task includes both fully virtual targets and targets that must be identified through real-world cues, ensuring that users must interpret both their physical environment and interface.} In this section, we describe the game's task area and overall structure.


\noindent\textbf{Game Structure.}
The structure of Pocket MonstARs is designed to \blue{approximate key demands of a typical warehouse inventory picking task} similar to that in our case study, while also incorporating all of \ProjectName's analytics and control capabilities. During the game, \blue{users assume the role of a warehouse worker within this abstracted task setting}, and are tasked with moving \ProjectName's map marker to direct the TurtleBot to specific locations in the task area so the robot can \blue{interact with real and virtual task targets (represented as ‘monsters’ in the game abstraction), doing this as quickly as possible. Several of these locations in each game variation require users to read instructions and use their real-world awareness to find their target, as required in both a real picking task and our case study where a worker must find a specific inventory location. In the game, monsters serve as targets requiring robot interaction rather than as literal warehouse entities. The  distinction between real and virtual targets allows us to separately measure performance on AR-guided interaction and real-world visibility-dependent components of the task.} Thus, \blue{the core characteristics of a common warehouse task are abstracted into the game design}, while allowing users to directly control the robotic collaborator in \blue{a gamified task environment designed to promote engagement while preserving task structure.}

\noindent\textbf{Task Area.}
The game task area, shown in Fig.~\ref{fig:teaser}(a), is a 2.4x2.4~m space constructed to resemble a section of a typical warehouse environment. While small relative to a full warehouse environment, the task area presents a greater challenge compared to a real-world large-scale environment, as AR situated analytics are more likely to obstruct important real-world targets. This makes it a strong test case for evaluating whether the \ObsPipeName can effectively mitigate AR obstructions. The area contains 16 identical 30.5x22.9x15.2~cm boxes distributed throughout that serve as inventory boxes, organized into rows akin to shelving lanes as found in a typical warehouse and our simulated case study. 12 boxes are labeled with randomly generated 9-character alphanumeric strings (e.g., ITMDXXIH3, ITMDAXIH3) on standard US letter-sized sheets of paper in 82~pt font. 4 boxes have objects detectable by the pretrained YOLO model~\cite{yolov5} atop them: 2 stuffed animals, a keyboard, and a potted plant. All labels appear visually identical but contain unique strings, with several differing by just 1 character. This use of labels and objects to identify specific boxes mirrors how distinct inventory items are organized in warehouses, where workers must locate and retrieve target items. The label similarity encourages users to read each label carefully, preventing reliance on memorized positions across game variations. \blue{These labels are therefore treated as critical task-relevant elements whose visibility must be preserved, motivating the use of text detection in the \ObsPipeName}.

\begin{figure*}[t]
  \centering
  \includegraphics[width=0.9\columnwidth]{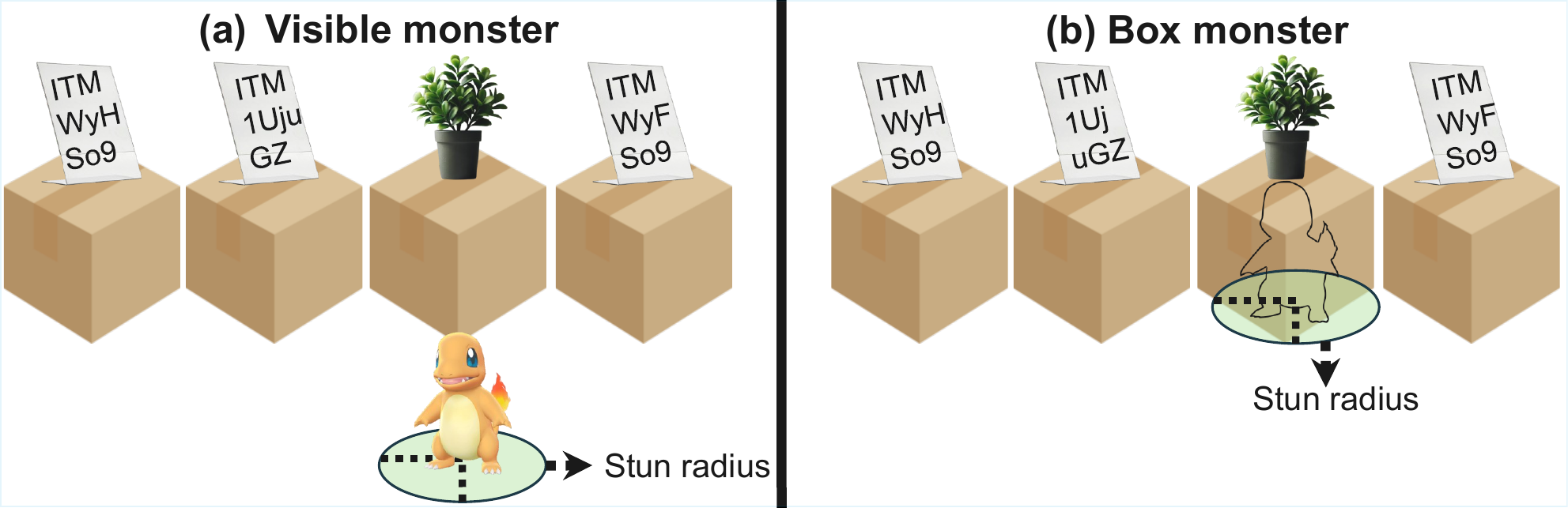}
  \caption{\blue{The two monster types in Pocket MonstARs: visible monsters and box-associated monsters identified using real-world cues.}}
  \Description{Two-panel diagram explaining the two monster target types in Pocket MonstARs. Panel (a), labeled “Visible monster,” shows several cardboard inventory boxes with printed labels and a potted plant on one box. In front of the boxes is a visible 3D virtual monster. A circular region on the floor around the monster is labeled “stun radius,” indicating the area the robot must reach in order to catch the monster. Panel (b), labeled “Box monster,” shows a similar row of labeled boxes, but no visible monster model is displayed. Instead, a dashed outline and circular stun radius appears inside one of the boxes, indicating that the monster is associated with a real-world box and must be found using real-world cues such as the box label or object. The figure contrasts targets that can be seen directly as virtual AR monsters with targets that require reading or recognizing physical inventory cues.}
  \label{fig:monstertypes}
\end{figure*}

\noindent\textbf{Task Design.}
To begin, users press a virtual start button using a "poke" gesture. While playing, users receive instructions on what their next target is from instruction text within the game. To complete the game, users must catch 15 monsters by moving the TurtleBot to each monster location and pressing a virtual stun button to make the TurtleBot stun the monster. A monster is considered stunned, and subsequently caught, if the TurtleBot's Euclidean separation distance to the monster as computed by the \SystemName is within 30~cm of the monster's center. This distance is referred to as the stun radius as shown in Fig.~\ref{fig:monstertypes}(a) and (b). Once the current monster is caught, the instruction text changes and directs users to the next monster. The game has 2 different types of monsters, as shown in Fig.~\ref{fig:monstertypes}, denoted \emph{visible monsters} and \emph{box monsters}. These two target types correspond to distinct aspects of warehouse picking tasks. \blue{Visible monsters represent tasks where the target location is directly identifiable, similar to retrieving items whose locations are already known. In contrast, box monsters represent tasks that require identifying a specific real-world item or location based on labels or objects, as in inventory picking where workers must locate a target box among visually similar options. This distinction ensures that the task includes both direct navigation and real-world search components, mirroring the range of perceptual demands present in warehouse scenarios.} 

10 visible monsters are represented by 3D models visible to the user, and appear at predefined locations with their stun radii centered at these locations. For this monster type, the instruction text would read "\emph{Catch the [monster name].}" Successful visible monster catches are indicated by the monster disappearing. \blue{To explicitly enforce reliance on real-world perception, we include a subset of targets that cannot be directly observed through AR content alone.} The 5 box monsters in the game are not directly visible, but are designated as being "inside" of 1 of the 16 inventory boxes within the space, \blue{requiring users to identify the correct real-world box using labels or objects}. These monsters also have predefined locations where their stun radii are centered, but these locations correspond to real-world boxes and no 3D model is rendered. For these monsters, the instruction text panel instructs users to "\emph{Save inventory box [box label string or box object]}", so they can direct the TurtleBot to drive near to that box and stun the monster inside. This requires visibility and awareness of the real-world task area. The stun procedure again involves the \SystemName computing the distance of the robot to the virtual monster inside the box, with a stun being successful if this distance is within the stun radius of the invisible monster. Successful box monster catches are indicated only by the instruction text changing. The visible monster location order and box selections are randomized once per game variation and remain consistent across all users for that variation. That is, every user experiences the same visible monster locations and boxes in any given variation as every other user that plays that variation. With this design, \emph{Pocket MonstARs requires users to have awareness of both virtual elements (e.g., \ProjectName analytics, virtual monsters) and real-world elements (e.g., box labels) to complete their task}, capturing the visibility and awareness demands of an AR-HRC warehouse picking task.

%% file: sec/study_design.tex
\section{User Study Design}
\label{sec:userstudydesign}
To evaluate ARTOO-DARTU, we conducted an IRB-approved 2 (Group: with vs. without the \ObsPipeName) $\times$ 3 (Trial: situated analytics variation) mixed-design user study with 34 participants using Pocket MonstARs. The independent variables were \textit{Group}, a between-subjects factor with two levels (\ObsPipeName active vs. inactive), and \textit{Trial}, a within-subjects factor with three levels: No Analytics (NA), Situated Analytics without arrows (SA), and Situated Analytics with arrows (SA+ARR). Each participant completed 3 variations of the game in a random order, each variation comprising a single trial. For all trials, users had access to the virtual map marker to control the TurtleBot, the instructions for their next target, and the stun button. 

\noindent\textbf{\ObsPipeName Group Splitting.}
To evaluate the \ObsPipeName in our game where real-world visibility is critical, participants were split evenly into 2 groups: a group with the \ObsPipeName active during all trials, and a group without it active. The \ObsPipeName group would observe virtual content obstructing their view of real-world labels and objects become transparent, while the non-\ObsPipeName group would not. This split enabled a between-subjects comparison of performance (e.g., completion and catch times) and survey responses in Sec.~\ref{sec:studyeval}.

\begin{figure*}[t]
  \centering
  \includegraphics[width=0.9\columnwidth]{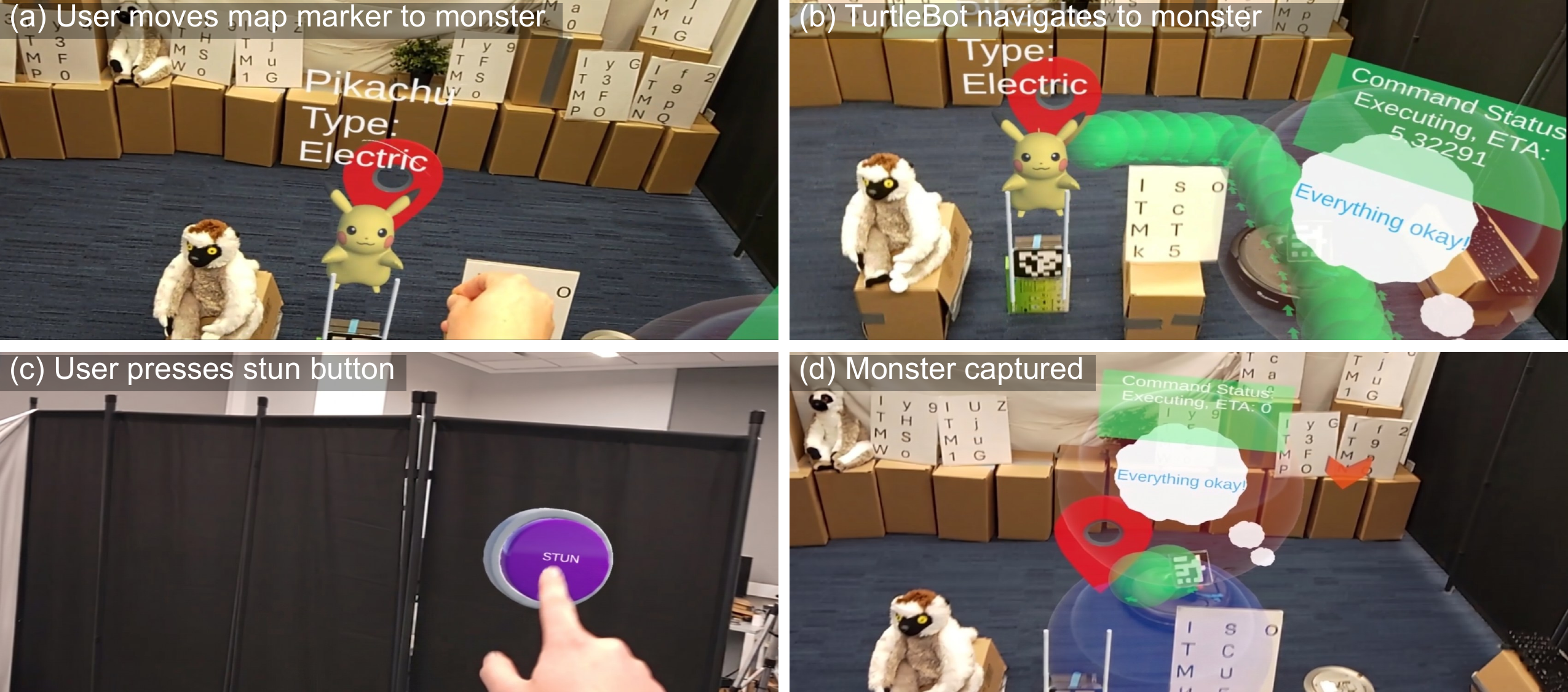}
  \caption{\blue{Process of catching a visible monster during the Pocket MonstARs SA+ARR trial with \ObsPipeName inactive. (a) A visible monster appears, and the user follows the instruction text by placing the map marker at the monster’s location. (b) The TurtleBot navigates to the map marker while trajectory indicators appear and the command status screen updates. (c) Once the TurtleBot reaches the destination, the user presses the stun button. (d) The monster is stunned, captured, and disappears.}}
  \Description{Four-panel sequence showing how a participant catches a visible monster during a Pocket MonstARs trial with situated analytics and arrows, with ODM inactive. In panel (a), a virtual monster appears in the task area near inventory boxes, and instruction text above it identifies the target as “Pikachu” with type “Electric.” A red AR map marker is placed at the monster’s location, showing the user’s command destination for the robot. In panel (b), the TurtleBot navigates toward the monster. Green AR trajectory indicators form a path on the floor, and a command status screen above the robot reports that the command is executing with an estimated time remaining. In panel (c), after the robot reaches the destination, the user presses a purple virtual “Stun” button with their finger. In panel (d), the monster has been captured and removed from the scene; the trajectory and AR robot analytics remain visible around the robot. The sequence illustrates the interaction loop: identify the target, place the robot destination marker, wait for robot navigation, press the stun button, and confirm the monster is captured.}
  \label{fig:trial-process}
\end{figure*}

\blue{\noindent\textbf{Visual Stimuli Trial Variations.}
During each trial, participants viewed the physical Pocket MonstARs task area (Fig.~\ref{fig:teaser}(a)) through the HoloLens 2 while seeing virtual task and robot-control elements overlaid in AR. The physical scene consisted of the task area containing the inventory boxes with labels and those with objects atop them. Across all trials, participants saw virtual instruction text above the back wall of the task area, indicating the next target, the map marker used to command the TurtleBot, and a virtual stun button used to capture the current monster as seen in Fig.~\ref{fig:trial-process}. For visible-monster targets, participants also saw a 3D virtual monster model at the target location (Fig.~\ref{fig:trial-process}(a) and (b)) and could direct the robot to that location by placing the map marker on the monster (Fig.~\ref{fig:trial-process}(a)). After receiving a command, the TurtleBot navigated toward the marker while participants could view the robot-control and analytics elements available in that trial, such as the trajectory indicators and command status screen in SA+ARR (Fig.~\ref{fig:trial-process}(b)). Once the TurtleBot reached the target, participants pressed the virtual stun button (Fig.~\ref{fig:trial-process}(c)); if the robot was within the monster's stun radius, the monster was captured and removed from the scene (Fig.~\ref{fig:trial-process}(d)). For box-monster targets, no monster model was rendered; instead, participants saw only the instruction text identifying a target box by its printed label or object. 

The three trial variants changed only the analytics available around this common task interface. In the \textit{No Analytics (NA) trial}, participants saw only the instruction text, map marker, stun button, and target monsters when applicable. In the \textit{Situated Analytics (SA) trial}, participants additionally saw the robot location indicator, thought bubble, command status screen, mini status indicators when applicable, and trajectory indicators showing the robot’s planned path. In the \textit{Situated Analytics with Arrows (SA+ARR) trial}, participants saw the same analytics plus orange arrow indicators above the current target inventory box, as in Fig.~\ref{fig:indicators}(c).}




\noindent\textbf{Analytics Feedback from Simulated Errors.}
To assess whether AR analytics helped reduce user frustration and maintain efficiency during unexpected robot behavior, each trial included 4 predefined moments simulating robot errors, during which the map marker disappeared and the TurtleBot ignored commands for 15~s. In the NA trial, participants had no feedback from the interface as to what was happening. In contrast, trials SA and SA+ARR accompanied these moments with a sound effect, the appearance of one of the mini indicators, and a personified message from the TurtleBot within the thought bubble. Once 15~s had elapsed, the marker reappeared, an "Everything okay!" message would display within the thought bubble, and a participant could again issue commands.

\subsection{Study Procedure and Participant Selection}
\label{sec:studyprocedure}
Upon arrival, participants read and signed a consent form and completed a pre-study survey. A researcher then introduced the game and gave a tutorial on using the map marker within the \SystemName. Each participant then completed 3 trials (avg. duration: 9.7~min), each followed by a post-trial survey and a 3~min break. After all trials, participants completed a post-experiment survey. 

We recruited 34 participants (mean age: 31.2 years, range: 19-67, 26\% female) from our university (N=19) and the broader community (N=15) to participate in the study. Each participant completed all 3 game trials. Due to technical issues, post-trial survey data was unavailable for 2 participants; performance data was available for all 34. Of the 34 participants, 3 use an AR headset more than once a week, 3 use a headset less than once a week, 13 had used a headset once or twice, and 15 had never used an AR headset.

\subsection{Data Collection}
\label{sec:datacollection}
\noindent\textbf{Survey Questions.}
After a trial, participants completed a survey comprised of a subset of questions from the Game Experience Questionnaire (GEQ)~\cite{gameexperiencequestion}, a subset from the System Usability Scale~\cite{sus} which we refer to as the Ease-of-Use (EOU) section, and Muir's Questionnaire for robot trust (MQ)~\cite{muirsquestion}. The GEQ assessed participants' opinions on the game by asking about their emotions while completing a trial, such as level of stimulation. The EOU gauged usability of \ProjectName in a trial. MQ was used to measure participants' trust in their collaborator in a trial. All responses were given on a 5-point Likert scale. Following all trials, participants ranked the 3 game variations by preference and were invited to explain their choices and provide open-ended feedback on the experience.

\noindent\textbf{Performance Data.}
During each trial, we collected data on overall catch times. For these catch times, the server API recorded the time from the moment a user was instructed to catch a certain monster to the moment that monster was caught. This data includes both visible and box monsters. In addition to overall times, we recorded catch durations for each visible and box monster separately using the same timing method, allowing for detailed analysis of performance by monster type.

%% file: sec/study_results.tex
\section{Results Analysis}
\label{sec:studyeval}
\begin{figure}[]
\centering
\includegraphics[width=\columnwidth]{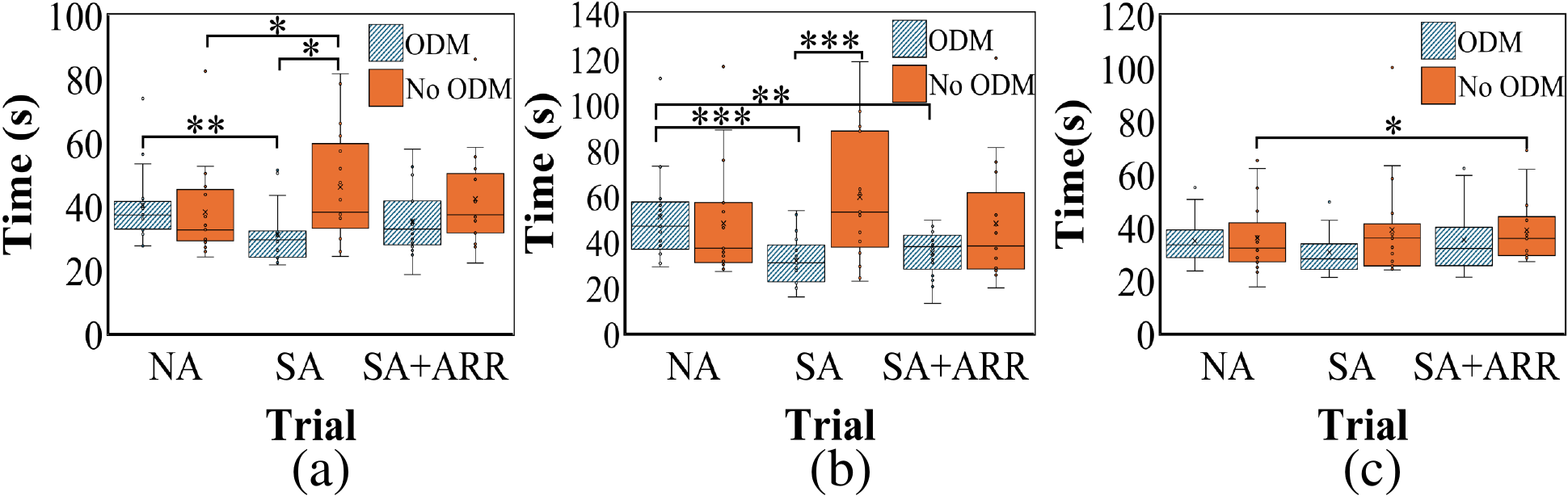}
\caption{Task performance data for all users of each group across all trials. (a): Average overall catch times. (b): Average box monster catch times. (c): Average visible monster catch times. (*): $p \leq .02$, (**): $p \leq .01$, (***): $p \leq .005$.}
\Description{Three box-and-whisker plots comparing task completion times across trial conditions for participants with ODM and without ODM. In all three plots, the y-axis is time in seconds, and the x-axis shows three trial types: NA, SA, and SA+ARR. Blue hatched boxes represent the ODM group, and orange boxes represent the no-ODM group. Panel (a) shows average overall catch time across all monsters. The ODM group is faster than the no-ODM group in the situated analytics condition, with a significant difference marked between groups; the ODM group also improves from NA to SA, while the no-ODM group becomes slower in SA. Panel (b) shows average box monster catch time, where the effect is stronger: the ODM group is substantially faster than the no-ODM group in SA, and the ODM group improves from NA to both SA and SA+ARR. This panel indicates that obstruction mitigation is especially helpful for targets requiring real-world label or object identification. Panel (c) shows average visible monster catch time. The differences are smaller than in the box-monster plot, and the main visible trend is that SA+ARR is associated with slower visible-monster catch times than SA. Significance brackets use one, two, or three asterisks to mark increasing levels of statistical significance. Overall, the figure shows that situated analytics improve performance most clearly when ODM is active, especially for real-world visibility-dependent box monster tasks.}
\label{userstudynumericalresults}
\end{figure}

This section discusses the effect of (1) \textbf{AR situated analytics} (denoted SA) and (2) \textbf{the \ObsPipeName} on task performance as measured by (1) average overall catch time, which is the average of all 15 monster catch times in each trial (2) average visible monster catch time, which is the average of the 10 catch times of the visible monsters in each trial, and (3) average box monster catch time, which is the average of the 5 catch times for box monsters only in each trial. While average catch time is a good indicator of overall task performance, we consider box monster catch times (which involved both real-world text and object identifications) to be different from visible monster catches (which involved only virtual content), and therefore report these 2 metrics separately. We report the results of these metrics in Fig.~\ref{userstudynumericalresults}. This section also discusses preference rankings from survey results.

We first conducted a mixed-design ANOVA, with 2 groups (with vs.~without \ObsPipeName) as the between-subjects factor and 3 trials (NA, SA, SA+ARR) as the within-subjects factor. Following ANOVA we conducted pairwise post-hoc tests, for which we chose the non-parametric Wilcoxon-Mann-Whitney U and Wilcoxon signed-rank tests due to the non-normality of the data. We set $\alpha = 0.05$ for individual tests and applied Bonferroni correction for multiple comparisons, and we report significance for each test according to this $\alpha$. For comparisons between trials SA and SA+ARR with an effect of \ObsPipeName and arrows, the significance threshold for each test was adjusted to 0.025 using the corrections.


\subsection{Objective Performance Metrics}

\noindent\textbf{Average Overall Catch Time.}
A mixed-design ANOVA revealed a significant main effect of the \ObsPipeName on average overall catch time [$F(1, 87) = 10.78$, $p = .001$, $n_p^2 = .13$], with the \textit{\ObsPipeName group performing significantly faster than the non-\ObsPipeName group} (31.8~s vs.~44.8~s, $p = .011$) with SA as presented in trial SA. In trial SA+ARR, the average catch time also decreased for the \ObsPipeName group (36.0~s vs.~41.8~s) compared to the non-\ObsPipeName group, but this difference was not significant ($p = .28$). A significant main effect of trial was also found [$F(2, 69)=3.54$, $p=.03$, $n_p^2=.09$], as post-hoc within-subjects tests in the \ObsPipeName group revealed a significant effect of SA with users catching monsters significantly faster in trial SA than in trial NA (31.8~s vs.~38.2~s, $p=.01$). Though not significant after correction, we found the \ObsPipeName group caught monsters slower in trial SA+ARR than in trial SA (36.0~s vs.~31.8~s, $p=.04$). The non-\ObsPipeName group showed a significant difference when comparing trial SA to trial NA (44.9~s vs.~35.5~s, $p=.02$), but not between trial SA+ARR and trial NA (41.8~s vs.~35.5~s, $p=.14$). These results suggest that while \textit{SA alone might decrease warehouse task efficiency if the SA obstruct the real-world, SA were effective in improving efficiency in the presence of the \ObsPipeName}. Arrows were another potential approach to improve task efficiency, but the effect was not significant, and even led to a slower catch time in the \ObsPipeName group.

\noindent\textbf{Average Box Monster Catch Time.}
The mixed-design ANOVA revealed an even more pronounced main effect of \ObsPipeName on average box monster catch time [$F(1, 76) = 15.65$, $p < .001$, $n_p^2 = .17$], with the \textit{\ObsPipeName group catching box monsters significantly faster than the non-\ObsPipeName group} (32.7~s vs.~57.9~s, $p = .002$) in trial SA, though there was no significance in trial SA+ARR (32.7~s vs.~49.4~s, $p=.22$). Within the \ObsPipeName group, users were significantly faster in trial SA than in trial NA (32.7~s vs. ~47.2~s, $p=.003$), and in trial SA+ARR than in trial NA (36.0~s vs. ~47.2~s, $p=.01$). In contrast, the non-\ObsPipeName group did not show significant differences on this metric, though users generally caught box monsters faster with the presence of arrows as in trial SA+ARR (49.4~s vs.~57.9~s). While the results generally corroborate the findings of overall catch time, the \textit{effect of the \ObsPipeName was found to be more pronounced for the box monster catch times, suggesting that the \ObsPipeName was especially effective in improving task efficiency in HRC tasks like inventory picking that involve intensive real-world searching}. However, this does not explain why users in the \ObsPipeName group performed slower in trial SA+ARR than in trial SA in overall catch times. 

\noindent\textbf{Average Visible Monster Catch Time.}
The mixed-design ANOVA revealed a significant main effect of trials on average visible monster catch time [$F(2, 68) = 3.78$, $p = .03$, $n_p^2 = .10$], with pairwise post-hoc tests finding a significant effect of SA and arrows in the non-\ObsPipeName group (36.0~s vs.~29.2~s, $p=0.02$), with users catching monsters significantly slower in trial SA+ARR than in trial NA. Though not significant, we also found SA alone to increase the average visible monster catch time in the non-\ObsPipeName group (38.4~s vs.~31.2~s, $p=.20$). Interestingly, while in the \ObsPipeName group SA alone was found to decrease average visible monster catch time (though not significant, 31.3~s vs.~33.7~s, $p=.20$), adding arrows conversely increased the average monster catch time (36.0~s vs.~31.3~s, $p=.05$). This suggests that no matter whether the \ObsPipeName is present, arrow indicators led to increased monster catch time. After careful examination of our system, this counterintuitive result can be explained by the fact that the arrows often pointed at a location ``over'' boxes in box monster catch tasks, which led to the users putting the navigation marker ``inside'' the box directly below the arrow indicator. These locations, however, were not reachable by the TurtleBot, which in turn led to the TurtleBot bumping into the box. While the user could still catch the box monster, in the successive visible monster catch task, the robot had to first back up and rotate before approaching the new destination, making the following catch unexpectedly slower. This is a limitation of the current design of the arrows, which we plan to address in future work.

\subsection{Subjective Feedback}
In this section we report survey results, including GEQ, EOU, MQ, ranking, and open responses. We treat the ordinal Likert scale as interval-level data and report the mean score of each question as answered by each user group in each trial. For ease of reporting, sometimes we combine the top 2 agreement categories (e.g., "extremely" and "fairly") into a single metric representing user endorsement of the statement, calculating the endorsement rate as the percentage of respondents selecting these combined categories. 


\noindent\textbf{GEQ Results.}
Among the ANOVA results of the GEQ questions, we found significant differences in responses to "\textit{I thought it was hard}" (main effect of \ObsPipeName [$F(1, 88) = 5.92$, $p=.017$, $n_p^2 = .08$]; interaction effect \ObsPipeName $\times$ Trial [$F(1, 61) = 10.99$, $p=.002$, $n_p^2=.13$]), "\textit{I was good at it}" (interaction effect [$F(1, 60) = 4.12$, $p=.05$, $n_p^2 = .06$]), "\textit{I was fast at reaching the game's targets}" (interaction effect [$F(1, 60) = 9.38$, $p=.003$, $n_p^2 = .11$]), "\textit{I thought it was fun}" (main effect of \ObsPipeName [$F(1, 86) = 6.97$, $p=.010$, $n_p^2 = 0.10$]), and "\textit{I felt competent}" (main effect of Trial [$F(2, 66) = 3.09$, $p=.05$, $n_p^2 = 0.09$]). 

Post-hoc tests revealed users in the \ObsPipeName group reported significantly lower endorsement rates of the experience being hard (1.3 vs.~2.1, $p=.002$) than those in the non-\ObsPipeName group, and significantly higher endorsement rates of ``being good at it'' (3.8 vs.~3.2, $p=.04$) and ``being fast'' (3.8 vs.~2.9, $p=.03$) in the AR trial. As for the effect of SA within the \textit{non-\ObsPipeName group}, users reported significantly lower endorsement rate of ``being good at it'' (3.2 vs.~3.7, $p=.05$) and ``felt competent'' (3.4 vs.~4.0, $p=.02$). For the effect of SA+\ObsPipeName, users reported significantly lower endorsement rates of ``thought it was hard'' (1.3 vs.~1.7, $p=.03$) and ``it was fun'' (3.9 vs.~4.3, $p=.04$), and significantly higher endorsement rates of ``being fast'' (3.8 vs.~3.2, $p=.02$). Regarding the effect of arrows, significantly lower endorsement rates of the experience being ``hard'' (1.4 vs.~2.1, $p=.01$) and higher endorsement rates of ``being fast'' (3.6 vs.~2.9, $p=.04$) were reported in the \textit{SA+ARR trial} compared to the \textit{SA trial}, only in the \textit{non-\ObsPipeName group}. Arrows did not significantly affect the \textit{\ObsPipeName group}, generally being outperformed by SA alone when the \ObsPipeName was present (though not significant, with higher endorsement of hardness (1.5 vs.~1.2, $p=.13$) and lower perceived efficiency (3.3 vs.~3.8, $p=.06$)).
Overall, the SA component of \ProjectName was found to enhance users' sense of competence and improve efficiency and simplicity when \textit{used in conjunction with the \ObsPipeName}, demonstrating how SA can significantly improve aspects of user experience in HRC warehouse tasks when a method is employed to keep the real-world unobstructed.

\noindent\textbf{EOU and MQ Results.} The mixed-design ANOVA did not find significant effects among the EOU and MQ questions. Therefore, we show endorsement results indicating the high usability and trustworthiness of the robot when using \ProjectName. Among all SA trials, users reported an endorsement rate of 71\% on ``\textit{I would like to use the system frequently},'' 85\%  on ``\textit{the system was easy to use},'' 90\% on ``\textit{the various functions in the system were well integrated},'' and 92\% on ``\textit{I would imagine that most people would learn to use the system very quickly}.'' Regarding trust, users reported an endorsement rate of 70\% on the overall trustworthiness of the robot, with 71\% endorsing the faith in the robot's ability to ``cope with similar situations in the future.'' Overall, \ProjectName was found to be highly usable and the robot trustworthy with SA in use, with users expressing a strong desire to use the system frequently and a high level of confidence in it. Unexpectedly, trust did not increase significantly in trials with SA, contrary to prior findings that AR generally boosts trust in HRC~\cite{SongVR2025VirtualUltrasound}. We suspect this is due to the TurtleBot’s familiar design, resembling a household robot vacuum cleaner, which participants may have had an established trust level with. We believe using less familiar, industrial-style robots may reveal stronger trust-related effects.

\noindent\textbf{Ranking and Open Responses.}
In the post-experiment survey, 83\% of participants ranked a trial with AR analytics as their favorite, with 65\% choosing SA+ARR and 18\% choosing SA. Open responses also reflected strong enthusiasm for using AR to interact with the TurtleBot: “\textit{It was a great experience. I definitely am convinced a robot along with AR can do a lot for operators},” and “\textit{I thought it was a really fun experience... I think this could be a big hit with a few changes made}.” One user simply stated, “\textit{I saw the future!}” Participants with the \ObsPipeName inactive frequently noted frustration with their view being obstructed; this feedback was not observed in the \ObsPipeName group, underscoring its effectiveness. These results suggest that \textit{users show strong positive sentiment for using situated analytics to collaborate with a robot in a warehouse task.}

%% file: sec/discussion_future_work.tex
\section{Discussion and Future Work}\label{sec:limitationsfw}
The results of our user study highlight the potential benefits of AR situated analytics in HRC warehouse scenarios when paired with our designed AR obstruction management method. We find that while situated analytics can significantly improve objective performance in our gamified warehouse task, this was only the case when real-world visibility was preserved using the \ObsPipeName; without obstruction mitigation, users experienced decreased efficiency. \ObsPipeName users also reported increased feelings of competence, efficiency, and simplicity compared to those without it. The overall response of users to the AR system was positive, with 83\% of users ranking a trial with situated analytics as their favorite and the majority of users stating the system was easy and enjoyable to use with analytics. These results suggest deploying AR-HRC systems with easily interpretable situated analytics into warehouse environments could enhance both user experience and task performance, but only when proper AR view management is applied.

\blue{

\subsection{Design Suggestions for AR Obstruction Detection and Mitigation}
While \ProjectName's \ObsPipeName is evaluated in an emulated warehouse task with gamified elements, our findings generalize to broader AR systems where virtual content coexists with task-relevant real-world elements. From our study results, we derive several AR design guidelines for future systems.

\noindent\textbf{Obstruction as a First-Class Concern.}
Our results regarding overall and box monster catch times in Sec.~\ref{sec:studyeval} suggest that obstruction management should not be treated as a secondary refinement, but as a prerequisite for effective AR assistance. Situated analytics only improved performance when obstructions were actively mitigated, indicating that visibility is foundational to the utility of AR content. Because obstruction can emerge dynamically as users, robots, and virtual elements move through the environment, mitigation must also operate in near real time. This favors efficient, local perception pipelines over heavy or cloud-dependent approaches, particularly in time-sensitive or privacy-constrained settings.

\noindent\textbf{Align Mitigation with Content Semantics.}
Mitigation strategies should be guided by the semantic role of virtual content. As described in Sec.~\ref{sec:systemdesign}, ARTOO-DARTU's situated analytics communicate the robot's position, internal state, movement status, and intended trajectory, making them spatially and semantically coupled to the physical robot. Because of this coupling, relocating these analytics during mitigation could weaken their relationship to the robotic collaborator. Accordingly, the \ObsPipeName uses visual modulation to restore real-world visibility while preserving spatial alignment. More broadly, AR elements tied to physical entities should preserve spatial coupling during mitigation, whereas less semantically grounded content may tolerate repositioning or suppression.

However, this approach becomes less suitable when AR content itself must be continuously monitored, as reduced visual prominence can impair readability. When alignment is less critical, alternative mitigation strategies may be more appropriate. Content can be displaced away from important real-world elements, scaled to reduce occlusion while maintaining legibility, or rendered with depth-aware occlusion behind real-world objects. More broadly, mitigation should be selected based on whether the task is primarily real-world-driven or AR-driven, rather than relying on a single universal strategy.

\noindent\textbf{Design for Density and Scale.}
As AR systems grow in complexity, designers must consider how increasing content density impacts obstruction. Our case study results in Sec.~\ref{sec:motivation} show that obstruction frequency rises with the number of active AR elements. In larger environments with more robots, obstruction is likely to become a persistent, system-level issue rather than an isolated event. Thus, at scale, reactive mitigation alone may be insufficient. Frequent or simultaneous obstructions can degrade both real-world visibility and the usability of AR content, particularly when transparency is employed. This suggests a need to move toward proactive management of content presentation. Systems should dynamically regulate what is shown based on task context, user focus, and content importance, using techniques such as context-aware filtering and adaptive abstraction. More broadly, while increased complexity enables richer analytics, these should be introduced selectively unless supported by robust content management strategies.
}
\blue{\subsection{Limitations and Future Work}}
There are several limitations in the user study presented in this work: only 1 robot collaborator was used, the study task and task area are a gamified emulation rather than a full warehouse deployment, our participants were not warehouse workers, \newblue{and the study did not include the range of safety-critical visual elements present in an actual warehouse.} \blue{Although Pocket MonstARs preserves key perceptual and coordination demands of warehouse HRC, including real-world target identification and obstruction opportunities, this abstraction was chosen to enable a controlled study in which all participants experienced comparable target locations, robot behaviors, and AR obstructions.} \newblue{However, real warehouses contain additional visual information that workers must continuously monitor, including moving equipment, warning signs, floor markings, stacked inventory, other workers, and other potential hazards. This is important because AR obstruction in such environments could affect not only task efficiency, but also worker safety. According to 2023 data from the US Occupational Safety and Health Administration (OSHA), at least 180 warehouse injuries occurred explicitly due to impairments of an employee's view of their environment~\cite{OSHAInjuryTracking_2023}. Naively implemented AR situated analytics applications could increase these types of injuries by obstructing a user's view of safety-relevant real-world information.} Future work should develop larger-scale studies with more collaborators, realistic inventory workflows, natural worker movement patterns, \newblue{and safety-relevant environmental features} to better understand how the \ObsPipeName performs when situated analytics more frequently obstruct the world. Investigations should also be performed in a real warehouse environment, recruiting warehouse workers as participants, thereby creating a study with stronger ecological and construct validity, \newblue{particularly with respect to tasks with safety-critical elements}. This direction is further motivated by prior work showing that environment and task selection can have a significant impact on AR studies, with participants performing better in a familiar environment~\cite{ZhangTVCG2024UserStudyEnvSelection}, suggesting a system like ours might have a more pronounced effect in a real warehouse scenario. \blue{Nevertheless, the present study isolates the central HRC trade-off addressed by our work: AR situated analytics can improve robot interaction, but only when the system also preserves visibility of task-relevant real-world information.}

\blue{
The \ObsPipeName also faces several key limitations. The pipeline relies on edge-based object and text detection models, and its effectiveness is therefore bounded by their accuracy in the target environment, which can vary with factors such as lighting conditions and viewing distance.
The models employed in this work are also general-purpose and not optimized for domain-specific features such as safety signage or specialized labels. In future work, we plan to explore alternative models and incorporate domain adaptation through model fine-tuning to better align detection with task-specific requirements.
The latency of the \ObsPipeName also remains a limitation. While acceptable under typical conditions, averaging 346~ms, delays between obstruction occurrence and mitigation can result in brief but repeated exposure to obstructed views. Future work will investigate addressing these challenges through improving system responsiveness, for example through on-device processing.
}

%% file: sec/conclusion.tex
\section{Conclusion}
\label{sec:conclusion}
This paper presents \ProjectName, an AR system that enables situated analytics and robot control during warehouse HRC tasks, featuring the \ObsPipeName, an obstruction mitigation pipeline designed specifically for AR-HRC. Our evaluation demonstrates that \ProjectName significantly increased user task efficiency, and its \ObsPipeName significantly reduced the impact of obstructive AR content, during the gamified warehouse task in our user study game Pocket MonstARs. Subjective feedback indicates that participants preferred the experience with analytics and the \ObsPipeName, reporting lower perceived difficulty and greater competence in the task.

\subsection*{Acknowledgments}
This work was supported in part by NSF grants CSR-2312760, CNS-2112562, and IIS-2231975, NSF CAREER Award IIS-2046072, NSF NAIAD Award 2332744, a CISCO Research Award, a Meta Research Award, Defense Advanced Research Projects Agency Young Faculty Award HR0011-24-1-0001, and the Army Research Laboratory under Cooperative Agreement Number W911NF-23-2-0224. The views and conclusions contained in this document are those of the authors and should not be interpreted as representing the official policies, either expressed or implied, of the Defense Advanced Research Projects Agency, the Army Research Laboratory, or the U.S. Government. This paper has been approved for public release; distribution is unlimited. No official endorsement should be inferred. The U.S. Government is authorized to reproduce and distribute reprints for Government purposes notwithstanding any copyright notation herein.